\documentclass{article}

\usepackage{arxiv}

\usepackage[utf8]{inputenc} 
\usepackage[T1]{fontenc}    
\usepackage{hyperref}       
\usepackage{url}            
\usepackage{booktabs}       
\usepackage{amsfonts}       
\usepackage{nicefrac}       
\usepackage{microtype}      
\usepackage{lipsum}
\usepackage{graphicx}
\usepackage{kotex}
\usepackage{subcaption}
\usepackage{rotating}

\title{THE TWO-SIDED MARKET NETWORK ANALYSIS\\
Based on Transfer Entropy \& Labelr}

\author{
  Seung Bin ~Baik \\
  DeepNatural Inc.\\
  \texttt{seungbin@deepnatural.ai} \\
}

\begin{document}
\maketitle
\begin{abstract}
This study more complex digital platforms in early stages in the two-sided market to produce powerful network effects. In this study, I use Transfer Entropy to look for super users who connect hominids in different networks to achieve higher network effects in the digital platform in the two-sided market, which has recently become more complex. And this study also aims to redefine the decision criteria of product managers by helping them define users with stronger network effects. With the development of technology, the structure of the industry is becoming more difficult to interpret and the complexity of business logic is increasing. This phenomenon is the biggest problem that makes it difficult for start-ups to challenge themselves. I hope this study will help product managers create new digital economic networks, enable them to make prioritized, data-driven decisions, and find users who can be the hub of the network even in small products.
\end{abstract}


\section{INTRODUCTION}
The two-sided market is is two groups of agents interact with each other via a common network platform and the value of participating in the network for agents in one group depends on the number of participants from the other group.[1] In this market, platform refer to goods and services that combine user groups.[2] Deep dive into this two-sided market, we can see that it consists of transaction parties affected by network effects and one or more intermediaries promoting their transactions.[3] These architectures of the network have become more complex in recent years, and it reminiscent of real-world complex networks. The network externality refers to the characteristics of the two-sided market that influence the outcome of the other group through the group's decision[4] is increasingly difficult to establish the causal relationship. Of course, there was uncertainty before, because the real drivers also affect it. But increasing complexity is causing more uncertainty. In particular, large platform companies, whose network effects have previously led to the growth of products through relationships among agents, have emphasized the network effects in homogeneous networks, while the business form of the recent platform is influenced by the influx of completely different groups.(e.g B2B2C) For this reason, it is difficult to find the highest-priority users to gain a competitive edge in the market and to measure the network effectiveness. Especially for crowd-sourcing platforms that do not provide communication functions, it is difficult to fully validate users' network effects and guess who is affecting external networks in the product. This study is based on a Transfer Entropy-based network analysis to solve these problems. And it aims to allow new clusters with certain criteria to influence different networks and to guess the connectivity of different external networks. And to best demonstrate these problems, we used data from the crowd-sourced data processing platform, "Labelr" provided by DeepNatural Inc. Labelr is a domestic service in Korea and provides users with web and mobile applications that can do data annotation for Machine Learning. 

Unlike other products, this product is divided into a  'Great user' group who was selected on have annotation experience and high inspection rates by internal professional staff. Therefore, Labelr is being chosen by customers for reason that high quality and speed of data annotation. Finally, I aim to define users(users with high network effects) who will focus on the product manager of the still-growing two-sided market platform and provide them with new decision criteria.

\section{TARGET MARKET AND PLATFORM}
Labelr, provided by DeepNatural Inc. is a platform that supplies the labor required to build artificial intelligence learning data with crowd-sourcing. The reason why crowd-sourcing platforms were selected for study is because of the structural features of two completely different networks interacting. For example, there is not much interaction in the user network, but this group of assistance to be rewarded, an influx of enterprise users(customers) who open the project is required. These enterprise users  do not choose to have vague and simple requirements like the number of users, outcomes. This structural specificity is consequently formed on other platforms or SNS, with the assistance group network disconnected without being formed internally. And enterprise users are also developed into networks between corporate representatives, making it difficult for product managers to focus on any single indicator in the platform.

To influence the effectiveness of these external networks and the network of enterprise customers, product managers have selected and managed separate groups of good assistants through internal evaluations. These tests have begun to create groups that receive huge rewards. However, these external networks are going through human hands, which could lead to another risk of loss of efficiency or intervention by stakeholders. In addition to the advantage of being able to analyze products that are rapidly growing in the artificial intelligence learning data market, which has clear market demand, the invisible hand has also worked greatly as much as the government's policy. These markets and products have the advantage of being similar to real society. 

\section{METHOD}
The data used in the study is the half-year data of 'Labelr' from July 2020 to December 2020. The reason why this data was selected is that was the period of growth, commonly referred to as J-curve. Also the inflow of B2B enterprise user was also increased, enhancing the production activities of the platform. It was also thought that data analysis of these growth periods would be more helpful in the decision-making of product managers who wanted to grow.\\
The data utilized credit related indicators. Credit is cash-like compensation for work in platform that user can withdraw anytime.

\begin{itemize}
\item \textit{Log data for the total amount of withdrawable credits remaining on the platform}
\item \textit{Total amount of credit paid to the user and the date of payment}
\item \textit{Total amount of credit withdrawn by the user and date of withdrawal}
\end{itemize}

We also utilized user-related values such as the following for an effective analysis of the network.

\begin{itemize}
\item \textit{New user and sign up date}
\item \textit{'Great user' who selected by internal professional staff and sign up date }
\item \textit{'Super user' calculated by the hypothesis and the sign up date}
\end{itemize}

Finally, we also used project data that opened in the 'Labelr', which can be interpreted as an indicator that reflects the actual demand for B2B.
We made a hypothesis that user networks in the platform will follow the power-law  in order to select 'Super user' calculated by the hypothesis. Agents that serve as hubs of the network will have the most impact.[5] To extract these hubs, The number of 'Super user' are calculated under the following conditions because the 'Labelr' cannot be connected to the network such as recommendation and relationship making.

\begin{enumerate}
\item Users with recent work logs and above average retention days
\item Users who are satisfying the conditions above and have more than average workload in their clusters
\item Users with an average pass rate of the first one or more task as a result of an expert examination of work results submitted by a cluster of users who are satisfied with the first condition
\end{enumerate}

We plan to use Transfer Entropy in this study to find out how these selected 'Super user' affect other drivers in the platform and to verify that 'Super user' will have more influence and network effects than 'Great user' selected by internal professional staff. Transfer Entropy is an indicator that quantifies the effect between two time series data.[6] This is a way to measure the direction of information as well as the absolute amount of information exchanged between the two systems. First, we obtain the slope of each element's variation to understand the shape of the change in the effect of the independent value(J) being the reference of the value of the two factors computed by the labeler's platform on the dependent value(I).

Below Formula 1 is an expression defined to determine the relationship between the previous sample of I and the previous sample of J.

\begin{equation}
T _{J\to I} = \sum p(i _{t+1} ,i _{t}^{(k)} ,j _{t}^{(l)} )\log {{p(i _{t+1} \vert i _{t}^{(k)} ,j _{t}^{(l)} )} \over {p(i _{t+1} \vert i _{t}^{(k)} )}}
\end{equation}

Finally, through the Transfer Entropy, I want to check the adequacy of the drivers and their effects that product managers should focus on the early stage two-sided market. Also I want to see if can draw a better smile graph with the Power User Curve, which allows us to see how network effect work.

\section{RESULT}
\subsection{Transfer Entropy}

The results of an analysis of time series data over six months based on transfer entropy through the key drivers of Labelr are shown in the Table 1 below.
In addition, the values in the table are replaced, converted to the final relative value(F), and listed in order, as shown in Table 1, and each transfer entropy unit is multiplied by 10,000 as arbitrary units(A.U.) and expressed to the third decimal place.

\begin{table}[htb]
  \mbox{}\hfill
    \caption{Labelr Transfer Entropy Result(A.U.)}
    \centering
    \begin{tabular}{c|c|c|c|c|c|c|c}
        \toprule
        \cmidrule(r){1-2}
       -     & User     & Great User     & Super User     & Credit     & Withdraw     & Remained Credit     & Project      \\
        \midrule
       User &  - &  72.156 &  0.001 & 40.853 & 0 & 42.557 & 132.956 \\
        Great User &  0 &  - &  0 & 29.303 & 25.447 & 30.544 & 95.355 \\
       Super User &  32.578 &  52.132 &  - & 29.499 & 25.618 & 30.749 & 95.994 \\
        Credit &  0 &  0 &  0.001 & - & 0 & 0 & 142.585 \\
        Withdraw &  5.84 &  0 &  0 & 5.288 & - & 0 & 17.213 \\
        Remained Credit &  0 &  0 &  0 & 1.516 & 1.317 & - & 4.934 \\
        Project &  0 &  0 &  0 & 0 & 0 & 0 & - \\
        \bottomrule
    \end{tabular}
\end{table}

For relative evaluation, the value of the correlation(T) resulting in Transfer Entropy is difficult to compare, and the effect of each value on the factor is divided by maximum value and reconstructed as follows with a score of 100 points.

\begin{equation}
F _{J\to I} = {{T _{J\to I}} \over {Max(T)}} \times 100
\end{equation}

\begin{table}[ht]
  \centering{}\hfill
    \caption{Labelr Impact Factor Ranking}
	\begin{tabular}{ll}
        \toprule
        \cmidrule(r){1-2}
                      & F     \\
                       \midrule
Credit to Project      & 100   \\
Super User to Projects & 67.32 \\
User to Great User        & 50.61 \\
User to Balance          & 29.85 \\
Super User to User        & 22.85 \\
Great User to Balance     & 21.42 \\
Great User to Credit      & 20.55 \\
Great User to Withdraw    & 17.85 \\
Withdraw to Users        & 4.1   \\
Remained Credit to Project      & 3.46  \\
Remained Credit to Withdraw      & 0.92  \\
\bottomrule
	\end{tabular}
\end{table}
As a result, the payment of credit has the highest impact on the project at 0.01425847AU, which is then converted back to the base (Max) to analyze the data. It was followed by 'user', 'super user', and 'great user'. Users who evaluated new users as experts in-house(Great User) also affected the increase in the number of automatically calculated best members(Super user). On the other hand, it was shown that the impact of the 'Great user' on other drivers was relatively low, especially the 34th lowest among 42 networks that led to an increase in total users. The important part of this study is what we focus on to increase the number of users and the number of projects being opened. So we need to look at the relationship between each driver more logically. Accordingly, when looked at the factors that have the most impact on the growth of users and the factors that have the most impact on the growth of projects. Naturally, for users with low impact but the barriers to participation are low, the experience of withdrawal and credit receipt (5.84 A.U. , 0 A.U. ) may be the biggest factor. But the 'Super user' had the most powerful impact with 32.578 A.U. on the increased users.
Project growth was most affected by the payment of credits at 142.575 A.U. on the opening of the project. As is generally said in the two-sided market, the next increase in users factor was 132.956 A.U., which had a powerful impact on the opening of the project. Overall, the project ranked 1st through 4th in the affected area, which could be the basis for a typical solution to the chicken and egg problem, which requires lower prices on two-sided market platforms or more inflows by first targeting subsidy sided users.

\subsection{Network Effect}
The Power User Curve is a measurement method of network effect. The stronger the network effect, the more the MAU user graph forms a smile.[7] The Power User Curve should be based on the original user's login, but Labelr did not have a logging system yet. So drawing the Power User Curve with a history of participating in the project. As a result, there are some differences in the degree of smile graphs, but it is expected that there will be no significant differences depending on the original purpose of the Power User Curve, which wants to see Network Effect.

\begin{figure}[ht]
     \centering
     \begin{subfigure}[b]{0.3\textwidth}
         \centering
         \includegraphics[width=\textwidth]{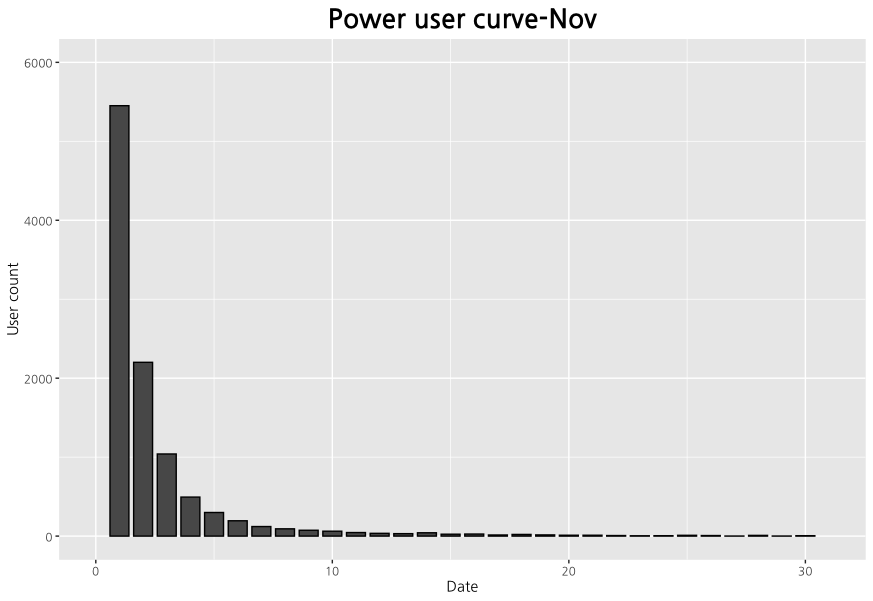}
         \caption{$User$}
         \label{fig:y equals x}
     \end{subfigure}
     \hfill
     \begin{subfigure}[b]{0.3\textwidth}
         \centering
         \includegraphics[width=\textwidth]{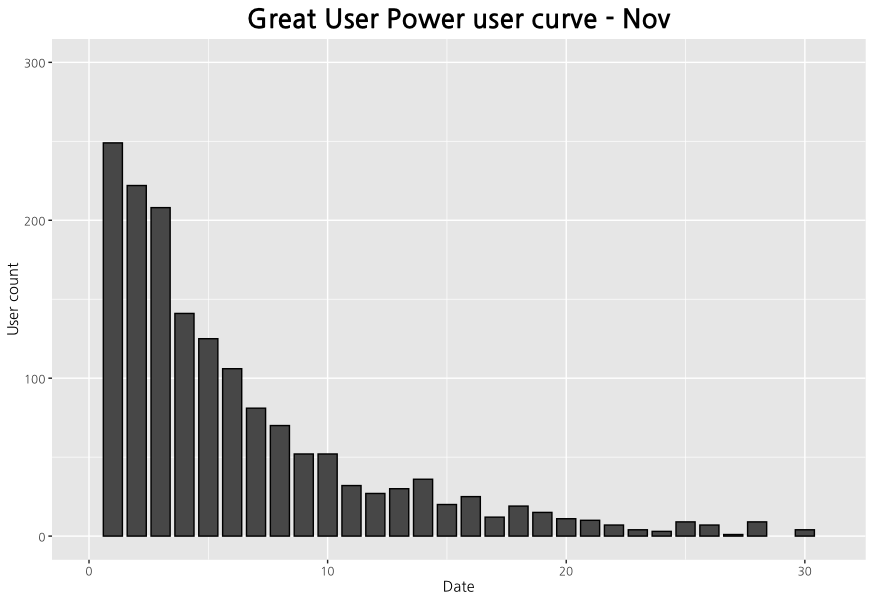}
         \caption{$Great User$}
         \label{fig:three sin x}
     \end{subfigure}
     \hfill
     \begin{subfigure}[b]{0.3\textwidth}
         \centering
         \includegraphics[width=\textwidth]{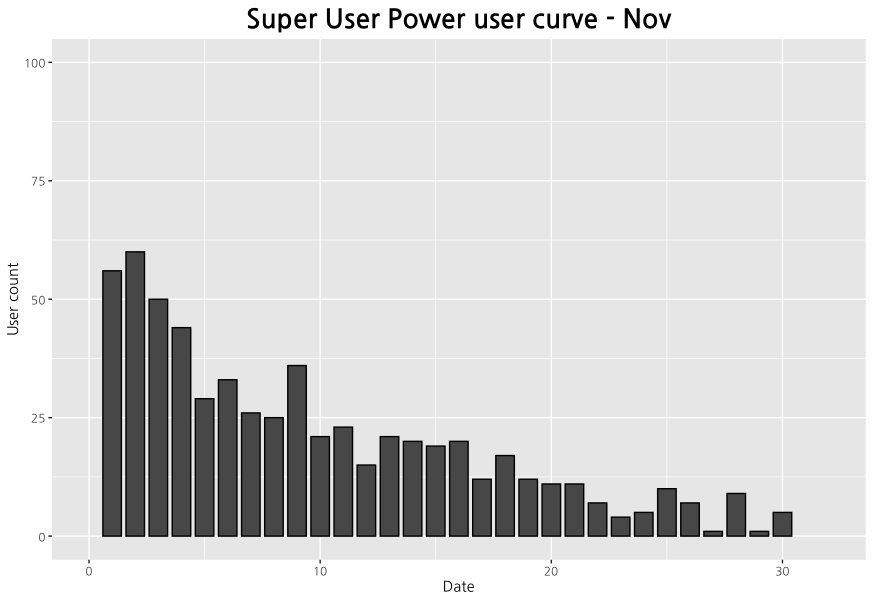}
         \caption{$Super User$}
         \label{fig:five over x}
     \end{subfigure}
        \caption{Power User Graph}
        \label{fig:three graphs}
\end{figure}

As the Power User Curve shows, 'Super user' consists of users with powerful network effects, and this metric provides various effects in most areas as One Metric That Matters(OMMTM) for digital platform services. And the Power User Curve shows that 'Super user' is most likely to become hubs in the platform's complexity network.

\section{CONCLUSION}

It was difficult to verify whether the 'Super user' selected through this study could be the hub of the power law. This is because there are no criteria for interaction and no user clusters interact in data users interacted outside the product. However, despite the nature of products that rely on different networks for indirect impacts, it is significant in that they have a similar size of impact to the 'Great user', which can only be selected through the internal testing. However, there is a limitation that it is hard to directly test the power law hypothesis in different networks and that it is hard to check whether the 'Super user' is in the hub. Furthermore, it is difficult to affirm that it is applicable as above on all platforms because there are not enough cases on other platforms.\\
To overcome the limitations of this study, I will analyze data from other products on the same basis to develop. And hope this study to help product managers make decisions that are creating new digital economic networks in their platform.

\section{ACKNOWLDEGEMENT}
Dedicate this paper to my family and colleagues who helped me conduct study and write a paper.\\
And also thanks DeepNatural Inc.(Team Labelr) for providing log data required for this study.

\clearpage
\bibliographystyle{unsrt}  


\end{document}